\documentclass[twocolumn,amsmath,amssymb]{revtex4}
\usepackage{amsfonts}
\usepackage{amsmath}

\begin{document}

\title{Supergravity in $(2+1)$ dimensions from $(3+1)$-dimensional Supergravity}

\author{P. Salgado}
 \email{pasalgad@udec.cl}
\author{F. Izaurieta}
 \email{fizaurie@udec.cl}
\author{E. Rodr\'{\i}guez}
 \email{edurodriguez@udec.cl}
\affiliation{Departamento de F\'{\i}sica, Universidad de Concepci\'{o}n,\\
Casilla 160-C, Concepci\'{o}n, Chile.}

\date{\today}

\begin{abstract}
In the context of the formalism proposed by Stelle-West and
Grignani-Nardelli, it is shown that Chern-Simons supergravity can
be consistently obtained as a dimensional reduction of
$(3+1)$-dimensional supergravity, when written as a gauge theory
of the Poincar\`{e} group. The
dimensional reductions are consistent with the gauge symmetries, mapping $%
(3+1)$-dimensional Poincar\`{e} supergroup gauge transformations onto $(2+1)$%
-dimensional Poincar\`{e} supergroup ones.

PACS number(s): 04.65. +e
\end{abstract}

\maketitle

\section{Introduction}

Supergravity in $(2+1)$ \cite{kay,achu} and in $(3+1)$
\cite{ps,ps1} dimensions can be formulated as a gauge theory of
the Poincar\`{e} superalgebra. The first-order formalism permits
one to write the three
dimensional supergravity as a Chern-Simons theory \cite{ban}, for which $%
(2+1)$-dimensional supergravity is a good theoretical laboratory for the
construction of a quantum theory \cite{witten}. Then it is interesting to
find a link between supergravities in $(2+1)$ and in $(3+1)$ dimensions.

The action for supergravity in $\left( 2+1\right) $-dimensions $S=\int
\left( \varepsilon _{abc}R^{ab}e^{c}+4\overline{\psi }D\psi \right) ,$ with $%
\psi $ a two component Majorana spinor, is invariant under Lorentz
rotations, Poincar\`{e} translations and supersymmetry transformations. The
dreibein $e_{\mu }^{a}$, the spin connection $\omega _{\mu }^{ab}$ and the
gravitino $\psi _{\mu }^{a}$ transform as components of a connection for the
super Poincar\`{e} group. This means that the supersymmetry algebra implied
by the corresponding supersymmetry transformations is the super Poincar\`{e}
algebra.

$(3+1)$-dimensional supergravity invariant under the Poincar\`{e}
supergroup is based on the supersymmetric extension of the
Stelle-West-Grignani-Nardelli formalism $(SWGN)$
\cite{stelle,grigna1,ps}. The fundamental idea of the formalism is
founded on the definition \cite{kobaya,stelle,ps} of the vierbein
$V^{A}$ and the gravitino $\Psi $, which involves the Goldstone
fields $\xi ^{A}$, $\chi
.$ In the supersymmetric extension of the $\left( SWGN\right) $-formalism: $%
(i)$ the vierbein $V^{A}$ is not identified with the component $e^{A}$ of
the gauge potential along the translation generators, but is given by
\begin{equation}
V^{A}=D\zeta ^{A}+e^{A}+i\left( 2\overline{\psi }+D\overline{\chi }\right)
\gamma ^{A}\chi ,  \label{cero}
\end{equation}%
$(ii)$ the gravitino field is not identified with the component $\psi $ of
the gauge potential along the supersymmetry generator, but is given by

\[
\overline{\Psi }=\overline{\psi }+D\overline{\chi }
\]

where $D\zeta ^{A}=d\zeta ^{A}+\omega ^{AB}\zeta _{B}$ , $D\chi =d\chi -%
\frac{1}{2}\omega ^{AB}\gamma _{AB}\chi $ and where $\omega ^{AB}$ is the
spin connection.

The purpose of the present work is to find the supersymmetric
extension of the successful formalism of refs.
\cite{grigna1,grigna2}. This means that, in the context of the
procedure of refs. \cite{grigna1,grigna2}, $(3+1)$-dimensional
supergravity can be dimensionally reduced to Chern-Simons
supergravity. This procedure can be used because both supergravity
in $(2+1)$ \cite{ban} and supergravity in $(3+1)$-dimensions
\cite{ps,ps1} can be formulated as theories genuinely invariant
under the Poincar\`{e} supergroup.

The paper is organized as follows: In sec.$II$, we shall review some aspects
of the Supersymmetric extension of the Stelle-West formalism and of
supergravity as a gauge theory of the Poincar\`{e} supergroup. The
dimensional reduction is carried out in sec.$III$ where the principal
features of the dimensional reduction process are presented. Section $IV$
concludes the work with brief comment.

\section{Supergravity invariant under the Poincar\`{e} group}

In this section we shall review some aspects of the Supersymmetric extension
of the Stelle-West formalism and of supergravity as a gauge theory of the
Poincar\`{e} group. The main point of this section is to display the
differences in the invariances of the supergravity action when different
definitions of vierbein are used.

\smallskip\

\subsection{Non-Linear realizations}

The non-linear realizations can be studied by the general method
developed in ref. \cite{callan,volkov}. Following these
references, we consider a Lie (super)group $G$ and a subgroup $H.$

Let us call $\left\{ {\bf V}_{i}\right\} _{i=1}^{n-d}$ the generators of $H$%
. We assume that the remaining generators $\left\{ {\bf A}_{l}\right\}
_{l=1}^{d}$ are chosen so that they form a representation of $H.$ In other
words, the commutator $\left[ {\bf V}_{i},{\bf A}_{l}\right] $ should be a
linear combination of ${\bf A}_{l}$ alone. A group element $g\in G$ can be
represented (uniquely) in the form
\begin{equation}
g=e^{\xi ^{l}{\bf A}_{l}}h  \label{sw1}
\end{equation}%
where $h$ is an element of $H.$ The $\xi ^{l}$ parametrize the coset space $%
G/H.$ We do not specify here the parametrization of $h$. One can define the
effect of a group element $g_{0}$ on the coset space by
\begin{equation}
g_{0}g=g_{0}(e^{\xi ^{l}{\bf A}_{l}}h)=e^{\xi ^{\prime l}{\bf A}%
_{l}}h^{\prime }  \label{sw2}
\end{equation}%
or
\begin{equation}
g_{0}e^{\xi ^{l}{\bf A}_{l}}=e^{\xi ^{\prime l}{\bf A}_{l}}h_{1}  \label{sw3}
\end{equation}%
where
\begin{equation}
\xi ^{\prime }=\xi ^{\prime }(g_{0},\xi )
\end{equation}%
\begin{equation}
h_{1}=h^{\prime }h^{-1}  \label{sw4}
\end{equation}%
\begin{equation}
h_{1}=h_{1}(g_{0},\xi ).
\end{equation}

If $g_0-1$ is infinitesimal, (\ref{sw3}) implies
\begin{equation}
e^{-\xi ^l{\bf A}_l}\left( g_0-1\right) e^{\xi ^l{\bf A}_l}-e^{-\xi ^l{\bf A}%
_l}\delta e^{\xi ^l{\bf A}_l}=h_1-1.  \label{sw5}
\end{equation}
The right-hand side of (\ref{sw5}) is a generator of $H.$

Let us first consider the case in which $g_0=h_0\in H.$ Then (\ref{sw3})
gives
\begin{equation}
e^{\xi ^{\prime l}{\bf A}_l}=h_0e^{\xi ^l{\bf A}_l}h_0^{-1}  \label{sw6}
\end{equation}
Since the $A^l$ form a representation of $H,$ this implies
\begin{equation}
h_1=h_0;\qquad h^{\prime }=h_0h.  \label{sw7}
\end{equation}

The transformation from $\xi $ to $\xi ^{\prime }$ given by (\ref{sw6}) is
linear. On the other hand, consider now
\begin{equation}
g_0=e^{\xi _0^l{\bf A}_l}.  \label{sw8}
\end{equation}
$.$ In this case (\ref{sw3}) becomes
\begin{equation}
e^{\xi _0^l{\bf A}_l}e^{\xi ^l{\bf A}_l}=e^{\xi ^{\prime l}{\bf A}_l}h.
\label{sw9}
\end{equation}
This is a non-linear inhomogeneous transformation on $\xi .$ The
infinitesimal form (\ref{sw5}) becomes
\begin{equation}
e^{-\xi ^l{\bf A}_l}\xi _0^i{\bf A}_ie^{\xi ^j{\bf A}_j}-e^{-\xi ^l{\bf A}%
_l}\delta e^{\xi ^i{\bf A}_i}=h_1-1.  \label{sw10}
\end{equation}
The left-hand side of this equation can be evaluated, using the algebra of
the group. Since the results must be a generator of $H$, one must set equal
to zero the coefficient of ${\bf A}_l.$ In this way one finds an equation
from which $\delta \xi ^i$ can be calculated.

The construction of a Lagrangian invariant under coordinate-dependent group
transformations requires the introduction of a set of gauge fields $a=a_\mu
^i{\bf A}_i$d$x^\mu ,$ $\rho =\rho _\mu ^i{\bf V}_i$d$x^\mu ,$ $p=p_\mu ^l%
{\bf A}_l$d$x^\mu $, $v=v_\mu ^i{\bf V}_i$d$x^\mu ,$ associated respectively
with the generators $V_i$ and $A_l$. Hence $\rho +a$ is the usual linear
connection for the gauge group $G,$ and the corresponding covariant
derivative is given by:
\begin{equation}
D=d+f(\rho +a)  \label{sw11}
\end{equation}
and its transformation law under $g\in G$ is
\begin{equation}
g:(\rho +a)\rightarrow (\rho ^{\prime }+a^{\prime })=\left[ g(\rho +a)g^{-1}-%
\frac 1f(dg)g^{-1}\right]  \label{sw12}
\end{equation}
where $f$ is a constant which, as it turns out, gives the strength of the
universal coupling of the gauge fields to all other fields.

We now consider the Lie algebra valued differential form \cite{callan}
\begin{equation}
e^{-\xi ^l{\bf A}_l}\left[ d+f(\rho +a)\right] e^{\xi ^l{\bf A}_l}=p+v.
\label{sw13}
\end{equation}
The transformation laws for the forms $p(\xi ,d\xi )$ and $v(\xi
,d\xi )$ are easily obtained. In fact, using
(\ref{sw8}),(\ref{sw9}) one finds \cite{zumino}
\begin{equation}
p^{\prime }=h_1p(h_1)^{-1}  \label{sw14}
\end{equation}
\begin{equation}
v^{\prime }=h_1v(h_1)^{-1}+h_1d(h_1)^{-1}.  \label{sw15}
\end{equation}

The equation (\ref{sw14}) shows that the differential forms $p(\xi ,d\xi )$
are transformed linearly by a group element of the form (\ref{sw8}). The
transformation law is the same as by an element of $H$, except that now this
group element $h_1(\xi _0,\xi )$ is a function of the variable $\xi $.
Therefore any expression constructed with $p(\xi ,d\xi )$ which is invariant
under the subgroup $H$ will be automatically invariant under the entire
group $G$, the elements of $H$ operating linearly on $\xi $, the remaining
elements non-linearly.

\smallskip\

\subsection{Supersymmetric Stelle-West Formalism}

The basic idea of the Stelle-West formalism is founded on the non-linear
realizations in anti de Sitter space \cite{stelle}. The supersymmetric
extension of this formalism \cite{ps1} is based in the non-linear
realizations of supersymmetry in anti de Sitter space \cite{zumino}. The
formalism consider as $G$ the graded Lie algebra

\[
\left[ P_A,P_B\right] =-im^2J_{AB}
\]
\[
\left[ J_{AB},P_C\right] =i\left( \eta _{AC}P_B-\eta _{BC}P_A\right)
\]
\[
\left[ J_{AB},J_{CD}\right] =i\left( \eta _{AC}J_{BD}-\eta _{BC}J_{AD}+\eta
_{BD}J_{AC}-\eta _{AD}J_{BC}\right)
\]
\[
\left[ J_{AB},Q_\alpha \right] =i\left( \gamma _{AB}\right) _{\alpha \beta
}Q_\beta
\]
\[
\left[ P_A,Q_\alpha \right] =-\frac i2m(\gamma _A)_{\alpha \beta }Q_\beta
\]
\begin{equation}
\left[ Q_\alpha ,\overline{Q}_\beta \right] =-2\left( \gamma ^A\right)
_{\alpha \beta }P_A-2m(\gamma ^{AB})_{\alpha \beta }J_{AB}  \label{ssu1}
\end{equation}
having as generators $Q_\alpha ,P_A$ and $M_{AB}$. It has as a subalgebra $H$
that of the de Sitter group $SO(3,2)$ with generators $P_A$ and $M_{AB}$.
This, in turn, has as subalgebra $L$ that of the Lorentz group $SO(3,1)$
with generators $M_{ab}.$ An element of $G$ can be uniquely represented in
the form
\begin{equation}
g=e^{\overline{\chi }Q}h=e^{\overline{\chi }Q}e^{-i\xi ^AP_A}l  \label{sw19}
\end{equation}
where $h\in H$ and $l\in L.$ On can define the effect of a group element $%
g_0 $ on the coset space $G/H$ by
\begin{equation}
g_0g=e^{\overline{\chi }^{\prime }Q}h^{\prime }=e^{\overline{\chi }^{\prime
}Q}e^{-i\xi ^{\prime A}P_A}l^{\prime }  \label{sw20}
\end{equation}
or
\begin{equation}
g_0e^{\overline{\chi }Q}=e^{\overline{\chi }^{\prime }Q}h_1  \label{sw21}
\end{equation}
\begin{equation}
h_1e^{-i\xi ^AP_A}=e^{-i\xi ^{\prime A}P_A}l_1  \label{sw22}
\end{equation}
\begin{equation}
l_1l=l^{\prime }.  \label{sw23}
\end{equation}
Clearly $h_1=h_1(g_0,\chi )$ and $l_1=l_1(g_0,\chi ,\xi ).$

If $g_{0}-1$ and $h_{1}-1$ are infinitesimal, (\ref{sw21}),(\ref{sw22})
imply
\begin{equation}
e^{-\overline{\chi }Q}\left( g_{0}-1\right) e^{\overline{\chi }Q}-e^{-%
\overline{\chi }Q}\delta e^{\overline{\chi }Q}=h_{1}-1  \label{sw24}
\end{equation}%
\begin{equation}
e^{i\xi ^{A}{\bf P}_{A}}\left( h_{1}-1\right) e^{-i\xi ^{A}{\bf P}%
_{A}}-e^{i\xi ^{A}{\bf P}_{A}}\delta e^{-i\xi ^{A}{\bf P}_{A}}=l_{1}-1.
\label{sw25}
\end{equation}%
We consider now the following cases: If $g_{0}=l_{0}\in L;$ (\ref{sw21}),(%
\ref{sw22}) give
\begin{equation}
e^{\overline{\chi }^{\prime }Q}=l_{0}e^{\overline{\chi }Q}l_{0}^{-1}
\label{sw26}
\end{equation}%
\begin{equation}
h_{1}=l_{1}=l_{0}  \label{sw27}
\end{equation}%
\begin{equation}
e^{-i\xi ^{\prime A}P_{A}}=l_{0}e^{-i\xi ^{A}P_{A}}l_{0}^{-1}.  \label{sw28}
\end{equation}%
Both $\chi $ and $\xi $ transform linearly. If, on the other hand, we know
only that $g_{0}=h_{0}\in H,$ in particular, if
\begin{equation}
g_{0}=e^{-i\rho ^{A}{\bf P}_{A}}  \label{sw29}
\end{equation}%
is a pseudo-translation, (\ref{sw21}) gives
\begin{equation}
e^{\overline{\chi }^{\prime }Q}=h_{0}e^{\overline{\chi }Q}h_{0}^{-1}
\label{sw30}
\end{equation}%
\begin{equation}
h_{1}=h_{0}  \label{sw31}
\end{equation}%
while (\ref{sw22}) gives
\begin{equation}
h_{0}e^{i\xi ^{A}{\bf P}_{a}}=e^{-i\xi ^{\prime A}P_{A}}l_{1}(h_{0},\xi ).
\label{sw32}
\end{equation}%
In this case $\chi $ transforms linearly, but the transformation law (\ref%
{sw32}) of $\xi $ under pseudo-translations is inhomogeneous and non-linear.
Infinitesimally
\begin{equation}
e^{i\xi ^{A}{\bf P}_{A}}\left( -i\rho ^{B}{\bf P}_{B}\right) e^{-i\xi ^{A}%
{\bf P}_{A}}-e^{i\xi ^{A}{\bf P}_{A}}\delta e^{-i\xi ^{A}{\bf P}%
_{A}}=l_{1}-1.  \label{sw33}
\end{equation}

Finally, if
\begin{equation}
g_0=e^{\overline{\varepsilon }Q}  \label{sw34}
\end{equation}
is a supersymmetry transformation, one must use (\ref{sw21}) and (\ref{sw22}%
) as they stand. Observe, however, that (\ref{sw22}) has the same form as (%
\ref{sw32}) except for the fact that $h_1$ is a function of $\chi $ while $%
h_0$ is not. Therefore, the transformation law for $\xi $ under a
supersymmetry transformation has the same form as that under a de Sitter
transformation but, with parameters which depend in a well defined way on $%
\chi .$

An explicit form for the transformation law of $\xi ^a$ under an
infinitesimal AdS boost can be obtained from (\ref{sw33}). The result is
\begin{equation}
\delta \xi ^A=-\rho ^A+\left( \frac{z\cosh z}{\sinh z}-1\right) \left( \rho
^A-\frac{\rho ^B\xi _B\xi ^A}{\xi ^2}\right)  \label{sw35}
\end{equation}
where $z=m\sqrt{(\xi ^a\xi _a)}=m\xi .$

The transformation of $\xi ^A$ under an infinitesimal Lorentz transformation
$l_0=e^{\frac i2\kappa ^{AB}J_{AB}}$ is
\begin{equation}
\delta \xi ^A=\kappa ^{AB}\xi _B  \label{sw36}
\end{equation}
and, under local supersymmetry transformation (\ref{sw34}), $\xi ^A$
transforms as
\[
\delta \xi ^A=-i\left( 1+\frac i6m\overline{\chi }\chi \right) \overline{%
\varepsilon }\gamma ^A\chi
\]
\[
+i\left( \frac{z\cosh z}{\sinh z}-1\right) \left( \delta _B^A-\frac{\xi
_B\xi ^A}{\xi ^2}\right) \left( 1+\frac i6m\overline{\chi }\chi \right)
\overline{\varepsilon }\gamma ^B\chi
\]
\begin{equation}
-2im\left( 1+\frac i6m\overline{\chi }\chi \right) \overline{\varepsilon }%
\gamma ^{AB}\chi \xi _B.  \label{sw37}
\end{equation}

Using (\ref{sw24}) with $g_0-1=\overline{\varepsilon }Q,$ one finds that
\begin{equation}
\delta \chi =\varepsilon -\frac i6m\left( 5\overline{\chi }\chi +\overline{%
\chi }\Gamma _A\chi \Gamma ^A\right) \varepsilon +\frac 19m^2\left(
\overline{\chi }\chi \right) \varepsilon  \label{sw38}
\end{equation}
\begin{equation}
h_1-1=\left( 1+\frac i6m\overline{\chi }\chi \right) \left( \overline{%
\varepsilon }\gamma ^A\chi P_A+m\overline{\varepsilon }\gamma ^{AB}\chi
J_{AB}\right) .
\end{equation}

Working in first order formalism, the gauge fields vierbein $e^A$, spin
connection $\omega ^{AB}$ and gravitino $\psi $ are treated as independent.
The key observation is that $(e^A,\omega ^{AB},\psi )$, considered as a
single entity, constitute a multiplet in the adjoint representation of the $%
AdS$ supergroup. That is, we can write:

\begin{equation}
A=\frac 12i\omega ^{AB}J_{AB}-ie^AP_A++\overline{\psi }Q  \label{ssu2}
\end{equation}
where $A$ is the gauge field of the AdS supergroup, $P_A,J_{AB},Q^\alpha $
being the generators of the AdS boosts. Then, based on these, we can define
the corresponding non-linear connections $(V^a,W^{ab},\Psi )$ from (\ref%
{sw13}):
\[
\frac 12iW^{AB}{\bf J}_{AB}-iV^A{\bf P}_A+\overline{\Psi }Q
\]
\begin{equation}
=e^{i\xi ^A{\bf P}_A}e^{-\overline{\chi }Q}\left[ d+\frac 12i\omega ^{AB}%
{\bf J}_{AB}-ie^A{\bf P}_A+\overline{\psi }Q\right] e^{\overline{\chi }%
Q}e^{-i\xi ^B{\bf P}_B}.  \label{sw39}
\end{equation}

If $G=OSp(1,4)$ and $H=SO(3,2),$ the gauge fields $V^A$ form a square $%
4\times 4$ matrix which is invertible and can be identified with the
vierbein fields. In the same way we have that $W^{AB}$ is a connection and
that $\overline{\Psi }$ can be identified with the Rarita-Schwinger field.
From (\ref{sw39}) one can obtain the fields$V^A,W^{AB},\Psi $ in terms of
the fields $e^A,\omega ^{AB},\psi .$ The results are given in equations $%
(81),(83)$ and $(84)$ of ref. \cite{ps1}.

The corresponding transformation laws for $V^a,W^{ab},\Psi $ can be obtained
from (\ref{sw14}),(\ref{sw15}). In fact, one can verify that, under the AdS
supergroup, the non-linear connections transform as:
\begin{equation}
\overline{\Psi }^{\prime }Q=h_1\left( \overline{\Psi }Q\right) (h_1)^{-1}
\label{sw40}
\end{equation}
\begin{equation}
-iV^{\prime a}{\bf P}_a=h_1\left( -iV^a{\bf P}_a\right) (h_1)^{-1}
\label{sw41}
\end{equation}
\begin{equation}
\frac 12iW^{\prime ab}{\bf J}_{ab}=h_1\left( \frac 12iW^{ab}{\bf J}%
_{ab}\right) (h_1)^{-1}+h_1d(h_1)^{-1}.  \label{sw42}
\end{equation}

The nonlinearity of the transformation with respect to the elements of $G/H$
means that the labels associated with the parts of the algebra of $G$ which
generate $G/H$ are no longer available as symmetry indices. In other words,
the symmetry has been spontaneously broken from $G$ to $H$. An irreducible
representation of $G$ will, in general, have several irreducible pieces with
respect to $H.$ Since, in constructing invariant actions, one only needs
index saturation with respect to the subgroup $H$, as far as the invariance
is concerned it is possible to select a subset of nonlinear fields with
respect to $G$, which form irreducible multiplets with respect to $H.$

\subsection{\bf Supergravity invariant under the AdS group}

Within the supersymmetric extension of the Stelle-West- formalism, the
action for supergravity with cosmological \cite{townsend} constant can be
rewritten as

\[
S=\int \varepsilon _{abcd}{\cal R}^{ab}V^cV^d+4\overline{\Psi }\gamma
_5\gamma _a{\cal D}\Psi V^a
\]
\begin{equation}
+2\alpha ^2\varepsilon _{abcd}V^aV^bV^cV^d+3\alpha \varepsilon _{abcd}%
\overline{\Psi }\gamma ^{ab}\Psi V^cV^d  \label{suads1}
\end{equation}
which is invariant under the supersymmetric extension of the AdS group. From
such equations we can see that the vierbein $V^a$ and the gravitino field
transform homogeneously according to the representation of the $AdS$
superalgebra but, with the nonlinear group element $h_1\in H.$

The corresponding equations of motion are obtained by varying the action
with respect to $\xi ^a,\chi ,e^a,\omega ^{ab},\psi $. The field equations
corresponding to the variation of the action with respect to $\xi ^a$ and $%
\chi $ are not independent equations. Following the same procedure of Ref.
\cite{salga1}, we find that equations of motion for supergravity genuinely
invariant under Super AdS are:
\begin{equation}
2\varepsilon _{abcd}\overline{{\cal R}}^{ab}V^c+4\overline{\Psi }\gamma
_5\gamma _d\rho  \label{suads5}
\end{equation}
\begin{equation}
\stackrel{\wedge }{{\cal T}}^{\ a}=0  \label{suads6}
\end{equation}
\begin{equation}
8\gamma _5\gamma _a\rho V^a-4\gamma _5\gamma _a\Psi \stackrel{\wedge }{{\cal %
T}}^{\ a}=0  \label{suads7}
\end{equation}

where
\begin{equation}
\stackrel{\wedge }{{\cal T}}^{\ a}={\cal T}^{\ a}-\frac i2\overline{\Psi }%
\gamma ^a\Psi  \label{suads8}
\end{equation}
\begin{equation}
\overline{{\cal R}}^{ab}={\cal R}^{ab}+4\alpha ^2V^aV^b+\alpha \overline{%
\Psi }\gamma ^{ab}\Psi =0  \label{suads9}
\end{equation}
\begin{equation}
\rho ={\cal D}\Psi -i\alpha \gamma ^a\Psi V^a.  \label{suads10}
\end{equation}

\subsection{Supergravity and the Poincar\`{e} group}

Taking the limit $m\rightarrow 0$ in equations $(24),$ $(73),$ $(75),$ $(76),
$ $(81),$ $(83)$ and $(84)$ one can see that: $(i)$ the superalgebra (\ref%
{ssu1}) take the form of the superalgebra of Poincar\`{e}
\[
\left[ P_{A},P_{B}\right] =0
\]%
\[
\left[ J_{AB},P_{C}\right] =i\left( \eta _{AC}P_{B}-\eta _{BC}P_{A}\right)
\]%
\[
\left[ J_{AB},J_{CD}\right] =i\left( \eta _{AC}J_{BD}-\eta _{BC}J_{AD}+\eta
_{BD}J_{AC}-\eta _{AD}J_{BC}\right)
\]%
\[
\left[ J_{AB},Q_{\alpha }\right] =i\left( \gamma _{AB}\right) _{\alpha \beta
}Q_{\beta }
\]%
\[
\left[ P_{A},Q_{\beta }\right] =0
\]%
\begin{equation}
\left[ Q_{\alpha },\overline{Q}_{\beta }\right] =-2\left( \gamma ^{A}\right)
_{\alpha \beta }P_{A}.  \label{su1}
\end{equation}

$(ii)$the transformation laws of $\xi ^{A}$ under an infinitesimal Poincar%
\`{e} translation, under an infinitesimal Lorentz transformation, and under
a local supersymmetry transformation are given respectively by

\begin{equation}
\delta \xi ^A=-\rho ^A  \label{poin0}
\end{equation}
\begin{equation}
\delta \xi ^A=\kappa ^{AB}\xi _B  \label{poin1}
\end{equation}
\begin{equation}
\delta \xi ^A=-i\overline{\varepsilon }\gamma ^A\chi ;  \label{poinc1}
\end{equation}
where $\rho ^A,$ $\kappa ^{AB}=-\kappa ^{BA}$ and $\varepsilon $ are the
infinitesimal parameters corresponding to Poincar\`{e} translations, Lorentz
rotations and supersymmetry respectively

$(iii)$ the transformation laws of $\chi $ under an infinitesimal Poincar%
\`{e} translation, under an infinitesimal Lorentz transformation, and under
a local supersymmetry transformation are given respectively by
\begin{equation}
\delta \chi =0  \label{poin2}
\end{equation}%
\begin{equation}
\delta \chi =\frac{1}{2}\kappa ^{AB}\gamma _{AB}\chi   \label{poin3}
\end{equation}%
\begin{equation}
\delta \chi =-\varepsilon .  \label{poin4}
\end{equation}

In this limit $G$ is the Poincar\`{e} supergroup and $H=SO(3,1);$ and the
fields vierbein $V^{A},$ the connection $W^{AB},$ and the Rarita-Schwinger
field $\overline{\Psi }$ are given by

\begin{equation}
V^A=e^A+D\zeta ^A+i\left( 2\overline{\psi }+D\overline{\chi }\right) \gamma
^A\chi  \label{poin5}
\end{equation}

\begin{equation}
W^{AB}=\omega ^{AB}  \label{poin6}
\end{equation}

\begin{equation}
\overline{\Psi }=\overline{\psi }+D\overline{\chi }  \label{poin7}
\end{equation}
where now

\begin{equation}
D=d+\omega .  \label{poin8}
\end{equation}
The corresponding components of the curvature two-form are now
\begin{equation}
{\cal T}\text{ }^A=DV^A  \label{poin9}
\end{equation}
\begin{equation}
R_B^A=d\omega _B^A+\omega _C^A\omega _B^C.  \label{poin10}
\end{equation}

\smallskip\

\section{Supergravity in $(2+1)$ from Supergravity in $(3+1)$}

\subsection{Supergravity in $(3+1)$}

The limit $m\rightarrow 0$ of the action \ref{suads1} is obviously the
action for $N=1$ Supergravity in $(3+1)$-dimensions:
\begin{equation}
S=\int \varepsilon _{ABCD}R^{AB}V^{C}V^{D}+4\overline{\Psi }\gamma
_{5}\gamma _{A}D\Psi V^{A}  \label{poin11}
\end{equation}%
which is genuinely invariant under the Poincar\`{e} group. In fact, $d=3+1$,
$N=1$ supergravity is based on the Poincar\`{e} supergroup, whose generators
$P_{A},J_{AB},Q^{\alpha }$ satisfy the Lie-superalgebra (\ref{su1}). Using
this algebra and the general form for gauge transformations on $A$
\begin{equation}
\delta A=-D\lambda =d\lambda -\left[ A,\lambda \right] ,  \label{cuatro}
\end{equation}%
with
\begin{equation}
\lambda =\frac{1}{2}i\kappa ^{AB}J_{AB}-i\rho ^{A}P_{A}+\overline{%
\varepsilon }Q,  \label{su4}
\end{equation}%
we obtain that $e^{A}$, $\omega ^{AB},$ and $\psi ,$ under local Lorentz
rotations, transform as
\begin{equation}
\delta e^{A}=\kappa _{B}^{A}e^{B};\quad \delta \omega ^{AB}=-D\kappa
^{AB};\quad \delta \psi =-\frac{1}{2}\kappa ^{AB}\gamma _{AB}\psi ;
\label{su5}
\end{equation}%
under local Poincar\`{e} translations, transform as
\begin{equation}
\delta e^{A}=D\rho ^{A};\quad \delta \omega ^{AB}=0\text{; \quad }\delta
\psi =0;  \label{su6}
\end{equation}%
and under local supersymmetry transformations, as
\begin{equation}
\delta e^{A}=-2i\overline{\varepsilon }\gamma ^{A}\psi ;\quad \delta \omega
^{AB}=0\text{; \quad }\delta \psi =D\varepsilon .  \label{su7}
\end{equation}

This means that the vierbein $V^{A}$ transforms, under the Poincar\`{e}
supergroup, as
\begin{equation}
\delta V^{A}=\kappa _{B}^{A}V^{B},  \label{ocho}
\end{equation}

The space-time supertorsion $\stackrel{\wedge }{{\cal T}\text{ }}^A$ is
given by
\begin{equation}
\stackrel{\wedge }{{\cal T}\text{ }}^A={\cal T}\ ^A-\frac 12\stackrel{\_}{%
\psi }\gamma ^A\psi ,  \label{once}
\end{equation}
where
\begin{equation}
{\cal T}\ ^A=DV^A.
\end{equation}

It is direct to verify that the action (\ref{poin11}) is invariant under (%
\ref{su5}),(\ref{su6}),(\ref{su7}), (\ref{poin0}),(\ref{poin1}),(\ref{poinc1}%
), (\ref{poin2}),(\ref{poin3}), (\ref{poin4}).

\subsection{Dimensional Reduction}

The dimensional reduction process, as well as the notation, is similar to
those used in refs. \cite{grigna1} and \cite{grigna2}. Latin indices $%
a,b,c,\cdot \cdot \cdot =0,1,2$ and capital latin indices $A,B,C,\cdot \cdot
\cdot =0,1,2,3$ denote $(2+1)$ and $(3+1)$ internal $($gauge$)$ indices
respectively. They are raised and lowered by the Minkowski metrics
\begin{equation}
\eta _{ab}=\left(
\begin{array}{lll}
-1 & 0 & 0 \\
0 & 1 & 0 \\
0 & 0 & 1%
\end{array}
\right) \text{ }
\end{equation}
and
\begin{equation}
\eta _{AB}=\left(
\begin{array}{llll}
-1 & 0 & 0 & 0 \\
0 & 1 & 0 & 0 \\
0 & 0 & 1 & 0 \\
0 & 0 & 0 & 1%
\end{array}
\right)
\end{equation}

In the dimensional reduction the first three values of $A,B,C,\cdot \cdot
\cdot $ will denote the corresponding $\left( 2+1\right) $ internal indices $%
a,b,c,\cdot \cdot \cdot ,$ i.e. $A=(a,3),$ $B=\left( b,3\right) ,$ $C=\left(
c,3\right) ,\cdot \cdot \cdot .$ We shall use the antisymmetric symbol $%
\varepsilon ^{ABCD\text{ }}$ with $\varepsilon ^{0123}=1$ and in $\left(
2+1\right) $-dimensions $\varepsilon ^{abc}=\varepsilon ^{abc3}$, so that $%
\varepsilon ^{012}=1.$

Following the procedure of ref. \cite{grigna1} we carried out a dimensional
reduction of the Poincar\`{e} generators of the $\left( 3+1\right) $%
-dimensional theory and, correspondingly, of the space-time dimensions that,
from the $\left( 3+1\right) $-dimensional action (\ref{poin11}) and the
algebra (\ref{su1}), lead to the $\left( 2+1\right) $-dimensional action.
With such reductions from the (3+1) gauge transformations (\ref{su5}),(\ref%
{su6}),(\ref{su7}), (\ref{poin0}),(\ref{poin1}),(\ref{poinc1}), (\ref{poin2}%
),(\ref{poin3}), (\ref{poin4}), we shall obtain the corresponding gauge
transformations in $\left( 2+1\right) $-dimensions.

The dimensional reduction leading from the $\left( 3+1\right)
$-dimensional supergravity theory to $(2+1)$-Chern-Simons
supergravity theory is given in the following Table (see
\cite{grigna1}):
\[
\begin{tabular}{||cc||}
\hline\hline
Dimensional & Reduction \\ \hline\hline
$\left( 3+1\right) $-dimensions & $\left( 2+1\right) $-dimensions \\
\hline\hline
$e^3$ & $dx^3$ \\ \hline\hline
$e^a$ & $e^a$ \\ \hline\hline
$\omega ^{ab}$ & $\omega ^{ab}$ \\ \hline\hline
$\omega ^{a3}$ & $0$ \\ \hline\hline
$\zeta ^a$ & $\zeta ^a$ \\ \hline\hline
$\zeta ^3$ & $0$ \\ \hline\hline
$\rho ^a$ & $\rho ^a$ \\ \hline\hline
$\rho ^3$ & $0$ \\ \hline\hline
$\kappa ^{ab}$ & $\kappa ^{ab}$ \\ \hline\hline
$\kappa ^{a3}$ & $0$ \\ \hline\hline
$\psi $ & $\psi $ \\ \hline\hline
$\gamma ^{abc}$ & $\gamma ^{abc}$ \\ \hline\hline
$\gamma ^3$ & 0 \\ \hline\hline
&  \\ \hline\hline
\end{tabular}
\]

where the $\gamma ^{\prime }$s with multiple indices are antisymmetrized
products of gamma matrices, which for $d$-dimensions satisfy the
relationship \cite{martin}
\begin{equation}
\gamma ^{i_1i_2\cdot \cdot \cdot \cdot \cdot \cdot i_k}=\alpha \varepsilon
^{i_1i_2\cdot \cdot \cdot \cdot \cdot \cdot i_d}\gamma _{i_{k+1}\cdot \cdot
\cdot \cdot \cdot \cdot i_d}\gamma ^{d+1}  \label{dieciseis}
\end{equation}
with
\begin{equation}
\alpha =\frac 1{\left( d-k\right) }\left( -1\right) ^{\frac 12k(k-1)+\frac 12%
d(d-1)}.
\end{equation}

It is direct to verify that the $\left( 3+1\right) $-gauge transformations (%
\ref{su5}),(\ref{su6}),(\ref{su7}), with the identifications of this Table
of dimensional reduction are mapped onto:
\begin{equation}
\delta \xi ^a=\kappa _{\text{ }b}^a\xi ^b;\delta e^a=\kappa _{\text{ }%
b}^ae^b;\text{ }\delta \omega ^{ab}=-D\kappa ^{ab};\delta \psi =-\frac 12%
\kappa ^{ab}\gamma _{ab}\psi ;
\end{equation}
\begin{equation}
\delta \xi ^a=-\rho ^a;\text{ }\delta e^a=D\rho ^a;\text{ }\delta \omega
^{ab}=0;\text{ }\delta \psi =0;
\end{equation}
\begin{equation}
\delta \xi ^a=-i\overline{\varepsilon }\gamma ^a\chi ;\text{ }\delta e^a=-2i%
\overline{\varepsilon }\gamma ^a\psi ;\text{ }\delta \omega ^{ab}=0;\text{ }%
\delta \psi =D\varepsilon ;
\end{equation}
i.e. onto the correct $\left( 2+1\right) $-dimensional gauge
transformations. In particular, the quantities that are set to a constant in
the Table consistently have vanishing gauge transformations. In the same way
we have
\begin{equation}
R^{AB}=\left(
\begin{array}{cc}
R^{ab} & R^{a3} \\
R^{3b} & R^{33}%
\end{array}
\right) =\left(
\begin{array}{cc}
R^{ab} & 0 \\
0 & 0%
\end{array}
\right)  \label{redu1}
\end{equation}
\begin{equation}
\omega ^{AB}=\left(
\begin{array}{cc}
\omega ^{ab} & \omega ^{a3} \\
\omega ^{3b} & \omega ^{33}%
\end{array}
\right) =\left(
\begin{array}{cc}
\omega ^{ab} & 0 \\
0 & 0%
\end{array}
\right)  \label{redu2}
\end{equation}
\begin{equation}
V^A=\left(
\begin{array}{c}
V^a \\
V^3%
\end{array}
\right) =\left(
\begin{array}{c}
e^a+D\xi ^a+i(2\overline{\psi }+D\overline{\chi })\gamma ^a\chi \\
dx^3%
\end{array}
\right)  \label{redu3}
\end{equation}
\begin{equation}
\Psi =\psi +D\chi  \label{redu4}
\end{equation}
where $D\xi ^a=d\xi ^a+\omega _b^a\xi ^b;$ $D\chi =d\chi -\frac 12\omega
^{ab}\gamma _{ab}\chi .$

From equation (\ref{dieciseis}) we see that, for $d=4$ and $k=3$,
\begin{equation}
\gamma ^{ABC}=-\varepsilon ^{ABCD}\gamma _D\gamma ^5  \label{veinte}
\end{equation}
which allows one to write the action for $\left( 3+1\right) $-dimensional
supergravity in the form
\begin{equation}
S^{4D}=\int \varepsilon _{ABCD}\left( R^{AB}V^CV^D+\frac 1{3!}\overline{\Psi
}\gamma ^{ABC}V^DD\Psi \right) .  \label{veintiuno}
\end{equation}
By substituting the content of the Table of dimensional reduction and (\ref%
{redu1}, \ref{redu2}) into the action (\ref{veintiuno}) one gets
\begin{equation}
S^{4D}=\int \left( 2\varepsilon _{abc3}R^{ab}V^c+\frac 4{3!}\varepsilon
_{abc3}\overline{\Psi }\gamma ^{abc}D\Psi \right) dx^3.  \label{redu5}
\end{equation}
Using (\ref{redu3}),(\ref{redu4}) and the identity $\gamma
_{ab}=-i\varepsilon _{abc}\gamma ^c$ we find that the first term is
\[
2\varepsilon _{abc3}R^{ab}V^cdx^3=(2\varepsilon
_{abc3}R^{ab}e^c+2\varepsilon _{abc3}R^{ab}D\xi ^c
\]
\begin{equation}
-4R^{ab}\overline{\psi }\gamma _{ab}\chi -2R^{ab}(D\overline{\chi })\gamma
_{ab}\chi )dx^3.  \label{redu6}
\end{equation}

Using (\ref{dieciseis}): $\gamma ^{abc}=-\varepsilon ^{abc}I$ and the
identities $DD\chi =\frac 12R^{ab}\gamma _{ab}\chi $; $\overline{\chi }%
\gamma _{ab}\psi =-\overline{\psi }\gamma _{ab}\chi $ we find that the
second term is
\[
\frac 4{3!}\varepsilon _{abc3}\overline{\Psi }\gamma ^{abc}D\Psi dx^3=(\frac %
4{3!}\varepsilon _{abc3}\overline{\psi }\gamma ^{abc}D\psi
\]
\begin{equation}
+4D(\overline{\chi }D\psi )+4R^{ab}\overline{\psi }\gamma _{ab}\chi
+2R^{ab}(D\overline{\chi })\gamma _{ab}\chi )dx^3.  \label{redu7}
\end{equation}

By substituting (\ref{redu6}) and (\ref{redu7}) in (\ref{redu5}) we obtain
\[
S^{4D}=\int (2\varepsilon _{abc3}R^{ab}e^c+\frac 4{3!}\varepsilon _{abc3}%
\overline{\psi }\gamma ^{abc}D\psi .
\]
\[
+2\varepsilon _{abc3}R^{ab}D\xi ^c+4D(\overline{\chi }D\psi ))dx^3.
\]

Using the Bianchi identity $DR^{ab}=0,\varepsilon _{abc}\varepsilon
^{abc}=-3!$ and (\ref{dieciseis}) with $d=3$ and $k=3,$ we find that the
action for $\left( 2+1\right) $- supergravity is given by
\begin{equation}
S^{4D}\longrightarrow S^{3D}=\int \varepsilon _{abc}R^{ab}e^c+4\overline{%
\psi }D\psi +surface\text{ }term.  \label{quince}
\end{equation}
which proves that the dimensional reduction from $\left( 3+1\right) $%
-dimensional supergravity to $\left( 2+1\right) $- supergravity is possible.

\section{ Comments}

We have shown that the successful formalism proposed in refs. \cite{grigna1}
and \cite{grigna2} can be extended to the supersymmetric case. That is, $%
(3+1)$-dimensional supergravity can be dimensionally reduced to supergravity
in $(2+1)$-dimensions following the method of refs.\cite{grigna1} and \cite%
{grigna2}.

Finally we can say that supergravity genuinely invariant under the Poincar%
\`{e} supergroup \cite{ps}, \cite{ps1} is a natural context to connect,
preserving the invariance under the Poincar\`{e} supergroup, such a theory
with $\left( 2+1\right) $-dimensional supergravity.

It is interesting to note that all the terms containing $\xi ^a,\chi $
disappear from the action and that $e^a,\psi $ can be interpreted as the
space-time dreibein and gravitino, and yet the theory is invariant under the
Poincar\`{e} supergroup, contrary to what happens in $\left( 3+1\right) $%
-dimensions. The absence of the $\xi ^a,\chi $ variables of (\ref{quince})
and the interpretation of $e^a,\omega ^{ab}$ and $\psi $ as gauge fields
makes of (\ref{quince}) an action that can be conceived as a Chern-Simons
three form.

\begin{acknowledgments}
This work was supported in part by Direcci\'{o}n de
Investigaci\'{o}n, Universidad de Concepci\'{o}n through Grant
N$_{o}$. 202.011.031-1.0. and in part by grant FONDECYT Projects
N$^{0}$ 1010485. (P.S) wishes to thank Herman Nicolai for the
hospitality in AEI in Golm bei Potsdam where part of this work was
done, and Deutscher Akademischer Austauschdienst (DAAD) for
financial support. (F.I) and (E.R) wish to thank Ricardo Troncoso
and Jorge Zanelli for their warm hospitality in Valdivia and for
many enlightening discussions. The authors are grateful to
Universidad de Concepci\'{o}n for partial support of the 2$^{nd}$
Dichato Cosmological Meeting, where this work was started.
\end{acknowledgments}

\end{document}